\documentclass[pre,aps,twocolumn,showpacs]{revtex4}
\usepackage{bm}
\usepackage{epsfig}

\usepackage{amsmath}
\usepackage{amssymb}

\newcommand{\ee}{\mathrm{e}}

\newcommand{\CC}{\mathcal{C}}
\newcommand{\BB}{\mathcal{B}}

\begin{document}

\title{Counting metastable states in a kinetically constrained model using a patch repetition analysis}
\author{Robert L. Jack}
\affiliation{Department of Physics, University of Bath, Bath, BA2 7AY, United Kingdom}

\begin{abstract}
We analyse metastable states in the East model, using a recently-proposed patch-repetition analysis based on time-averaged density profiles.
The results reveal a hierarchy of states
of varying lifetimes, consistent with previous studies in which the metastable states were identified and used to explain
the glassy dynamics of the model.  
We establish a mapping between these states and configurations
of systems of hard rods, which allows us to analyse both typical and atypical metastable states.  We discuss connections
between the complexity of metastable states and large-deviation functions of dynamical quantities, both in the context of the
East model and more generally in glassy systems.  
\end{abstract}

\pacs{64.70.Q-, 05.40-a}

\maketitle

\section{Introduction}

As supercooled liquids approach their glass transitions, their viscosities (and relaxation times)
increase rapidly~\cite{nagel-review,deb-review}.  
Despite a very considerable body of theoretical work, there remains
no consensus as to how this observation should best be explained.  In some theoretical pictures,
relaxation takes place via ``excitations''~\cite{GC-pnas-2003,GC-annrev-2010} (or ``soft spots''~\cite{manning11}) 
that become increasing rare at low temperatures, 
slowing down the dynamics.  Alternatively, one may imagine that the system evolves on a rough potential energy 
surface, and tends to become trapped in deep minima as the temperature is lowered~\cite{goldstein69,heuer08}.  Or perhaps, the diversity
(entropy) of disordered states decreases so strongly at low temperatures that transitions between distinct
states require large-scale rearrangements that are necessarily very slow~\cite{KTW,BB-ktw-2004}.  

In order to refine these qualitative pictures, theoretical developments are complemented by 
computer simulations of glassy fluids. However, the practical task of identifying objects like ``excitations'' or ``free energy minima'' 
in simulation is a difficult one (see however~\cite{manning11,keys-prx2011}).  
A particular case in point is the idea of \emph{metastable states}.  It is very natural to describe glassy dynamics in 
terms of rare transitions between such states, which are 
are also well-defined in some classes of mean-field model~\cite{tap,cav-review}.  
In computer simulation, identifying distinct metastable states 
is possible in small systems~\cite{heuer08}, but the generalisation to extended (large) systems is more 
difficult  (see also~\cite{biroli-kurchan}).  This is an important obstacle when attempting to test
theoretical pictures based on mean-field models.

Recently, Kurchan and Levine~\cite{KL} proposed a \emph{patch-repetition analysis}
whereby metastable states may be identified and characterised,
using computer simulations of large systems.  
The method has two main components: a time-averaging procedure that associates
metastable states of lifetime $\tau$ with well-defined density profiles; and a counting procedure based on finite `patches' of
the large system.  
This work 
provides a method (or thought-experiment) by which metastable states may be defined in finite-dimensional systems.
Recent numerical studies based on this method~\cite{camma-patch,sausset11} have focussed on the counting procedure, showing that this can
indeed yield an entropy associated with the number of states in the system.  

Here, we apply the patch-repetition analysis
to the East model~\cite{Jackle91,SE}.  This is a kinetically constrained model~\cite{Ritort-Sollich,GST-kcm} whose
thermodynamic properties (and potential energy landscape) are trivial, but whose dynamics are nevertheless complex,
and have striking similarities with molecular glass-formers~\cite{nef2005,ecg-fit2009,keys-east2013}.  We find that the patch-repetition
analysis can be used to identify and characterise metastable states in this model, despite its trivial
potential energy surface.  However, the metastable states that are revealed absolutely require the time-averaging
process described in~\cite{KL}: in particular, there are states with a very broad spectrum of lifetimes, and the
results of the patch-repetition analysis depend on the lifetime of the states under analysis.  
In this sense, the patch-repetition analysis is distinct from recent studies that aim to characterise amorphous order
through analyses where a system is constrained (pinned) or biased to remain close to a typical reference
configuration~\cite{biroli-pin,berthier-kob-pin,jack-plaq-pin,berthier-silvio} -- these analyses are purely 
static and have no dependence on the dynamical rules by which
the system evolves.  Such calculations therefore give trivial results for the East model, 
in contrast to the patch-repetition analysis.

We describe our models and methods in Sec.~\ref{sec:model}
before showing numerical results in Sec.~\ref{sec:results}.  
We also identify a mapping between these states and configurations of an
ideal gas of hard rods, where the size of the rods depends on the lifetime of the states.  
In Sec.~\ref{sec:discuss}, we discuss the implications of our results for studies of metastable states
in general, and 
we also discuss connections between that analysis and recent work on large
deviations of dynamical quantities in glassy systems~\cite{Merolle,kcm-transition,hedges09,jack-rom09}.
Finally, we summarise our conclusions in Sec.~\ref{sec:conc}.

\section{Model, methods, and metastable states}
\label{sec:model}

\subsection{Model}

The East model~\cite{Jackle91,SE} 
consists of $L$ binary spins $n_i=0,1$, with $i=1\dots L$, and periodic boundaries.  We refer
to spins with $n_i=1$ as `up' and those with $n_i=0$ as `down'.  
The notation $\CC=(n_1,n_2,\dots,n_L)$ indicates a configuration of the system.  
The key feature of
the model is that spin $i$ may flip only if spin $i-1$ is up.  If this \emph{kinetic constraint}~\cite{Ritort-Sollich,GST-kcm}
is satisfied, spin $i$ flips from $0$ to $1$ with rate $c$ and from $1$ to $0$ with rate $1-c$.
We take $c/(1-c) = \ee^{-\beta}$ where $\beta$ is the inverse temperature,
so small $c$ corresponds
to low temperature.  
The rates for spin flips
obey detailed balance, so that the probability of
 configuration  $\CC$ at equilibrium is simply $p^0(\CC) = \ee^{-\beta\sum_i n_i}/Z$,
where $Z=(1+\ee^{-\beta})^L$ is the partition function.
At equilibrium, one has $\langle n_i \rangle=c$, and the regime of interest is small $c$ (low temperature).

\begin{figure}
\includegraphics[width=8cm]{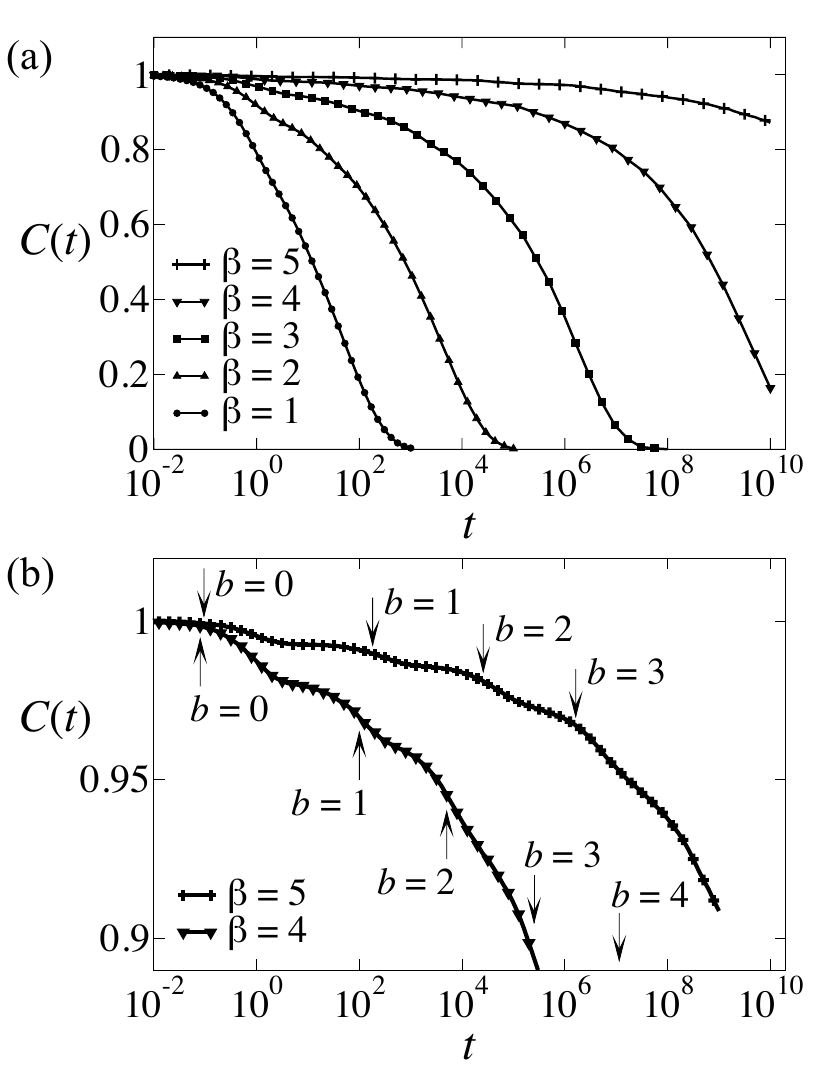}
\caption{Correlation function $C(t)$ in the East model.  (a)~The inverse temperature is varied from $\beta=1$ to
$\beta=5$, showing a dramatic increase in relaxation time.  (b)~At the lowest temperatures, plateaus are
apparent during the early stages of relaxation (where $C(t)$ is large).  Arrows indicate the times
for which data is shown in later figures.  The behaviour at these times is representative of regimes $c^{1-b} \ll t \ll c^{-b}$:
the rationale for choosing the specific values of $t$ is discussed in the main text.}
\label{fig:cor}
\end{figure}

Despite the trivial form of $p^0(\CC)$, the kinetic constraint in this model leads to complex co-operative
dynamics at small $c$.  In particular, the relaxation time of the model diverges in a super-Arrhenius fashion,
as $\tau_0 \sim \ee^{\beta^2/(2\ln2)}$~\cite{SE,East-gap}.  The rapid increase in relaxation time is illustrated 
in Fig.~\ref{fig:cor}a where we show 
\begin{equation}
C(t) = \frac{ \langle n_i(t) n_i(0) \rangle - \langle n_i \rangle^2 }{ c(1-c) } ,
\end{equation}
for various temperatures.
The origin of the increasing time scale is a large set of metastable states, arranged
hierarchically~\cite{SE}: different states have different lifetimes which scale as $c^{-b}$.  Here and throughout
$b$ is an integer which we use to classify the various relevant time scales in the system. 
Evidence for the separated timescales is shown in Fig.~\ref{fig:cor}b: the correlation function decays 
via a sequence of plateaus that become more clearly separated as $c$ is reduced.

\subsection{Metastable states}

To identify metastable states, we perform a time average over the spins, defining
\begin{equation}
  \overline{n}_{it} = (1/t) \int_0^t \mathrm{d}t' n_i(t'),
\end{equation}
and a time-averaged spin profile 
\begin{equation}
\overline{\CC}_t = (\overline{n}_{1t},\overline{n}_{2t},\dots,\overline{n}_{Lt}).
\end{equation}
The key observation of~\cite{KL} is that if the system has metastable states indexed by $\alpha=1,2,\dots$, 
then each state is associated
with a profile $\overline{\CC}^\alpha$.  Further, if $t$ is much larger than the intrastate relaxation times, but small
compared to their lifetimes, then the observed
profile $\overline{\CC}_t$ will almost surely be close to one of the $\overline{\CC}^\alpha$.  Hence, the statistical
properties of the (observable) profiles $\overline{\CC}_t$ can be used to infer the properties of the metastable
states in the system.

This situation, of many metastable states each associated with a profile $\overline{\CC}^\alpha$,
holds very accurately in the East model.  
As discussed by Sollich and Evans~\cite{SE} (see also~\cite{Aldous02}), motion on a time scale $t\ll c^{-b} \ll \tau_0$ is restricted to domains of size
$2^{b-1}$, with each domain being immediately to the right of a long-lived up spin.  Also, if $c^{1-b} \ll t \ll c^{-b}$ then
spins within such domains typically flip many times within time $t$, while other spins are unlikely to flip at all.  The result
of this large number of flips is that $\overline{n}_{it}\to \langle n_i\rangle = c$ if spin $i$ is within a mobile domain.
Sollich and Evans~\cite{SE} used the term ``superdomain'' to describe these mobile domains.  A metastable state with lifetime of
order $c^{-b}$ can be identified by specifying the position of its superdomains.

In terms of Fig.~\ref{fig:cor}b, the limit $c^{1-b} \ll t \ll c^{-b}$ corresponds to choosing a time within a plateau of the 
correlation function.  However, when using numerical results to gain information about metastable states, 
we find that saturating the limit $t\gg c^{1-b}$ is more important
than $t\ll c^{-b}$: in the following we focus on the time points indicated by arrows in Fig.~\ref{fig:cor}b,
which correspond to $c^{1-b} \ll t \lesssim c^{-b}$.  However, 
our results are similar if we use smaller times (as long as $t \gg c^{1-b}$).

\begin{figure}
\begin{center}
\includegraphics[width=8.5cm]{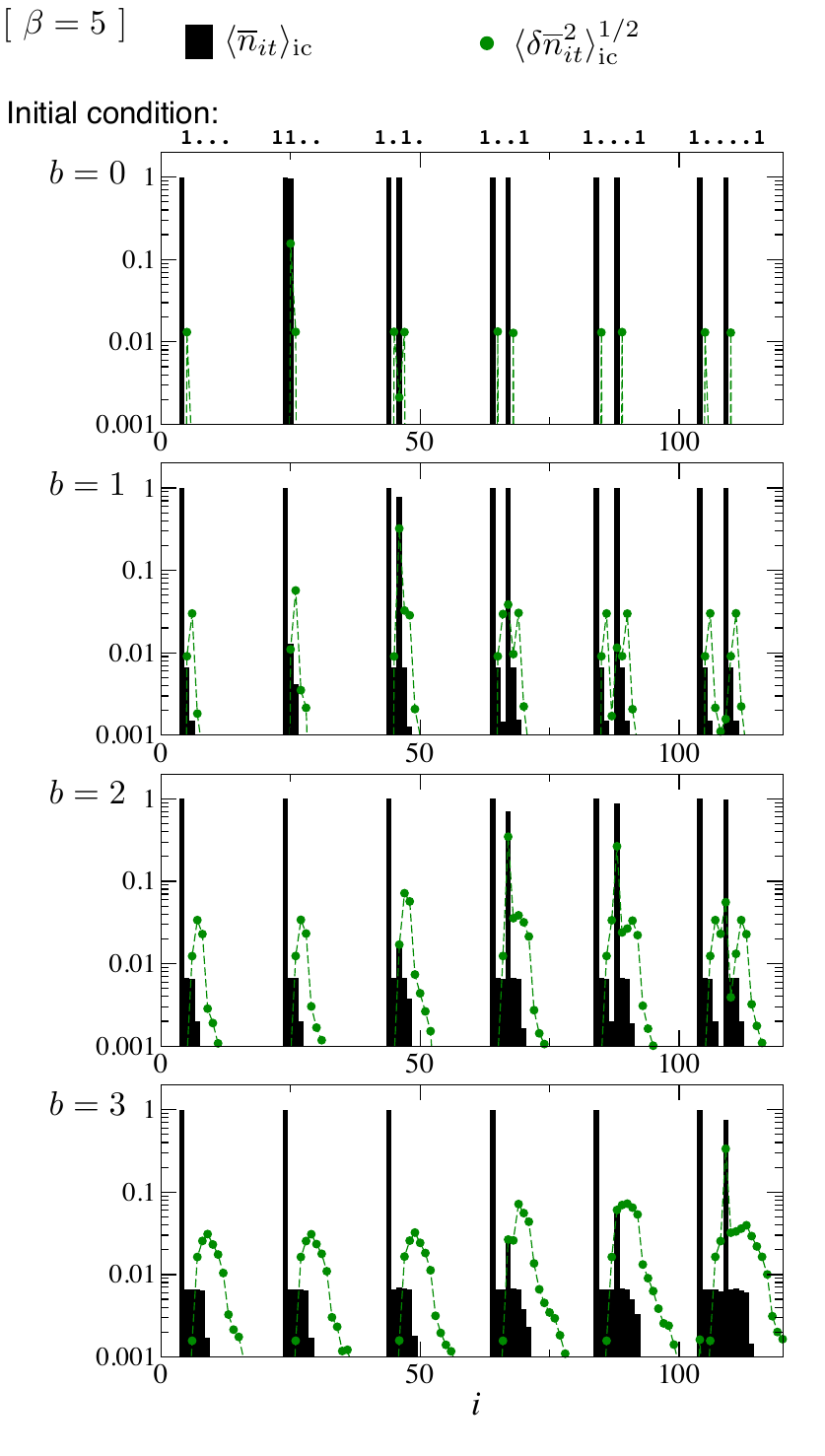}
\end{center}
\caption{
  Time averaged density profiles $\overline{\CC}_t$ at $\beta=5$ (so $c \approx 0.0067$).
  Averaging over many trajectories from the same initial condition, we show the mean profile
  $\langle \overline{n}_{it} \rangle_{\rm ic}$ and its standard deviation 
  $\langle (\delta \overline{n}_{it})^2 \rangle_{\rm ic}$.  The times correspond with the arrows in Fig.~\ref{fig:cor}b,
  and are typical of the regimes $c^{1-b}\ll t \ll c^{-b}$.  The approximate positions of the up spins in the initial condition are 
  shown at the top of the figure ({\tt 1}s indicate up spins and dots indicate selected down spins).  
  Typically, $\overline{n}_{it}$ is either close to $1$ (i.e., 
  $\langle \overline{n}_{it} \rangle_{\rm ic}\approx 1$ with $\langle (\delta \overline{n}_{it})^2 \rangle_{\rm ic}\ll1$
  or $\overline{n}_{it}$ is small (i.e., $\langle \overline{n}_{it} \rangle_{\rm ic}\ll 1$ 
  with $\langle (\delta \overline{n}_{it})^2 \rangle_{\rm ic}\ll1$).  The separation between these two cases
  motivates the definition of the (coarse-grained) $\tilde{n}_{it} = 0,1$).
}
\label{fig:prof}
\end{figure}

To illustrate this behaviour,  
we have simulated trajectories of the East model,
starting from a particular initial condition that helps to reveal the physical processes at work.
Fig.~\ref{fig:prof} shows the time-averaged spins $\langle \overline{n}_{it} \rangle_{\rm ic}$ and their variances
$\langle (\delta \overline{n}_{it})^2 \rangle_{\rm ic}$, 
where the subscript
``ic'' indicates that the average was taken with a fixed initial condition (in contrast to other averages in this 
work which are conventional averages over equilibrium trajectories).
We emphasise the following three
points:
\begin{itemize}
\item All of the $\langle \overline{n}_{it} \rangle_{\rm ic}$ are either close to $1$ or of order $c\ll1$.
Also, the standard deviation $\langle (\delta \overline{n}_{it})^2 \rangle_{\rm ic}^{1/2} \ll 1$ for all spins.
Thus, for this initial condition and for each of these times $t$,
one almost certainly finds a profile $\overline{\CC}_t$ that is close to a reference
profile $\overline{\CC}^\alpha_t = \langle \overline{n}_{it} \rangle_{\rm ic}$.  
Each reference profile (one for each value of $t$ shown) can therefore be identified with a metastable state
that is stable on a time scale $t$.
\item For $b\geq1$, any spins with $ \langle \overline{n}_{it} \rangle_{\rm ic} \approx 1$ are separated by at least 
$2^{b-1}$ spins with $\langle \overline{n}_{it} \rangle \ll 1$.  If two up spins are closer than this in the initial condition then
the rightmost of them is within the mobile domain associated with the leftmost one.  In that case, the rightmost
spin ($n_j$) tends to flip many times and $\overline{n}_{jt}$ converges to a value close to $c$.
\item For this initial condition (and these time scales), 
the system relaxes independently in each of 6 independent regions.  This means that the dynamical
evolution of separate patches of the system can be treated independently. 
\end{itemize}

\begin{figure}
\begin{center}
\includegraphics[width=8.5cm]{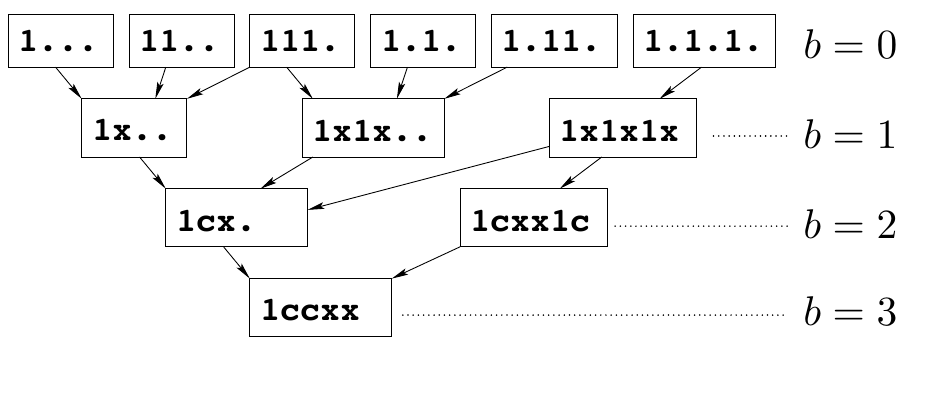}
\end{center}
\caption{Schematic illustration of time evolution among metastable states.  Each box represents a state, labelled
 by its characteristic local density profile, and a 
  value of $b$ that indicates its lifetime (of order $c^{-b}$). 
To describe the profile we use a notation suggested by Fig.~\ref{fig:prof}, based on the relative sizes 
of $\langle \overline{n}_{it}\rangle_{\rm ic}$ and $\langle \delta\overline{n}_{it}^2\rangle_{\rm ic}^{1/2}$.
We distinguish four cases: (i)~a {\tt 1} represents a spin with $\overline{n}_{it}\approx 1$ (up to a tolerance much less than $1$); 
(ii)~a dot
represents a spin with $\overline{n}_{it}\ll c$;
(iii)~a {\tt c} represents a spin with $\overline{n}_{it}\approx c$ (up to a tolerance much less than $c$);~(iv) an {\tt x} represents a spin with $\overline{n}_{it}\sim c$ (that is, a typical value $c$ with a tolerance much less than $1$ but bigger
than $c$).  The {\tt 1} corresponds to $\tilde{n}_{it}=1$ while {\tt x}, {\tt c} and dot all correspond to 
$\tilde{n}_{it}=0$.  Arrows indicate how a profile on one time scale evolves into a different profile after averaging
over a longer time scale.
Some states have more than one outward arrow, indicating that they may evolve into one of several possible profiles,
so they are on a borderline between the various (more stable) states into which they may evolve.
}
\label{fig:hier}
\end{figure}

These three points establish the central requirements for our analysis of metastable states.
The first shows that the metastable states in this system are well-defined, while the second
shows that metastable states with different lifetimes have different internal structures.  The third point
establishes that one may decompose the behaviour of the system into independent regions, at least for
this initial condition.  Building on this point, Fig.~\ref{fig:hier} 
shows how the local behaviour evolves with time, for various different initial conditions 
(the predictions shown follow from the superspin analysis of~\cite{SE} and
are also consistent with our numerical results).  We note that
while some initial conditions belong
to a single metastable state (and always yield the same time-averaged profile $\overline{\CC}_t$), 
there are other initial conditions that exist on the border
between states (and therefore may yield one of several profiles).
For these cases, it will not be true that $\langle (\delta \overline{n}_{it})^2 \rangle_{\rm ic}^{1/2} \ll 1$,
but it is true that for typical trajectories, $\overline{\CC}_t \approx \overline{\CC}_t^\alpha$ for
some metastable state $\alpha$.  

In general, the analysis of Kurchan and Laloux~\cite{laloux} indicates that typical configurations
of any large finite-dimensional system always lie on a border between states.  For the East model, this
result follows from the observation that the local configuration ${\tt 111...}$ will occur many times in a large system:
in the regions where this occurs, the system may relax into one of two averaged profiles [either ${\tt 1x1x..}$ or ${\tt 1xx...}$,  
(see Fig.~\ref{fig:hier})].
Hence a typical configuration of the system
cannot be associated to a single averaged profile.  However,
a key insight from~\cite{KL} is that while it is not possible to establish a one-to-one mapping between
initial conditions and metastable states, it is possible to establish such a mapping between 
profiles $\overline{\CC}_t$ and metastable states. 

\subsection{Methods for counting metastable states}

We now turn to the patch-repetition analysis proposed by Kurchan and Levine~\cite{KL} as a method
for analysis of metastable states.  We wish to consider the 
probability distribution of profiles $\overline{\CC}_t$, since these
are in one-to-one correspondence with metastable states.  In general, the averages
$\overline{n}_{it}$ have continuous values, and a tolerance $\epsilon$ is required~\cite{KL} in order to
identify if a state $\overline{\CC}_t$ is ``close'' to a reference profile $\overline{\CC}_t^\alpha$.
However, in the East model, we have $\overline{n}_{it}\approx 0,1$ (recall Fig.~\ref{fig:prof}) so we 
define
\begin{equation}
  \tilde{n}_{it} =  \Theta( \overline{n}_{i,t} - a) ,
  \label{equ:thresh}
\end{equation}
 where $\Theta(x)$ is a step function, and the threshold $a=1/2$ 
(results depend weakly on this threshold).
We then obtain a binary profile 
\begin{equation}
  \tilde{\CC}_t = (\tilde{n}_{1t},\tilde{n}_{2t},\dots,\tilde{n}_{Lt}),
\end{equation} 
and we consider the statistical properties of these profiles, as a proxy for the $\overline{\CC}_t$.

To analyse the distribution over the $\tilde{\CC}_t$, we use
the patch-repetition analysis~\cite{KL}.  To this end, 
consider a patch of the system of size $\ell$, for example 
$\BB_i^\ell=(\tilde{n}_{it},\tilde{n}_{i+1,t},\dots,\tilde{n}_{i+\ell-1,t})$.
Then, for a large system, one evaluates
\begin{equation}
S_{\ell,t} = -\sum_{\BB^\ell} \left[ n(\BB^\ell)/L \right] \log\left[ n(\BB^\ell)/L \right]
\label{equ:S-patch}
\end{equation}
where $n(\BB^\ell)$ is the number of occurences of patch $\BB^\ell$ in $\tilde{\CC}_t$, and the sum runs
over all patches of size $\ell$ that appear in $\tilde{\CC}_t$.
(Clearly $\sum_{\BB^\ell} n(\BB^\ell) = L$ since the total number of patches is $L$.)  

As long as correlations in the system are of finite range, one may define
the entropy density associated with the profiles $\tilde{\CC}_t$ as
\begin{equation}
  {\cal S}_t = \lim_{\ell\to\infty} S_{\ell,t}/\ell .
\end{equation}
Since we expect a one-to-one correspondence between profiles $\tilde{\CC}_t$ and 
metastable states of lifetime at least $t$, then we can identify ${\cal S}_t$ as the entropy density
associated with these metastable states.  The (extensive) quantity $L{\cal S}_t$ is sometimes called 
the ``complexity''~\cite{KL,tap,cav-review} (although we emphasise that we are working with states of fixed finite
lifetime, not the infinitely long-lived states that exist in mean-field models).

\section{Patch-repetition analysis and dynamical correlations}
\label{sec:results}

\begin{figure}
\includegraphics[width=8.5cm]{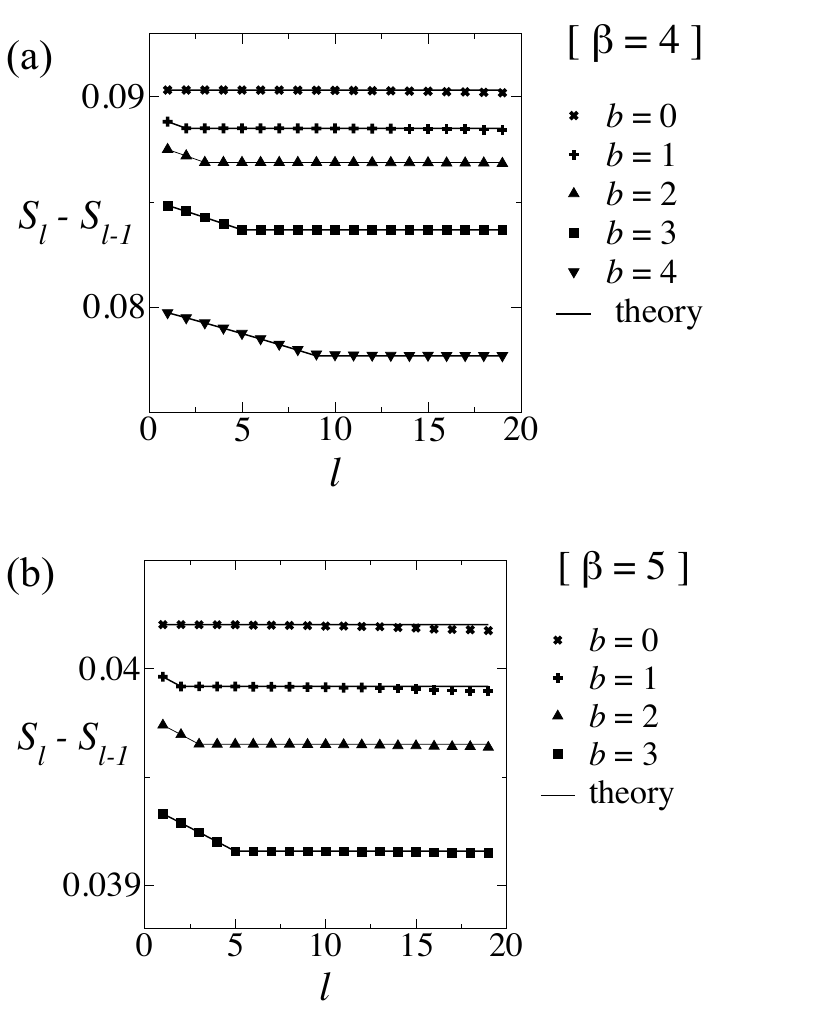}
\caption{Patch entropies $S_{\ell,t}$ (points) associated with the density profiles $\tilde{n}_{it}$,
evaluated at the times indicated in Fig.~\ref{fig:cor}b and for (a) $\beta=4$ and (b) $\beta=5$.  The `theory' lines are the predictions
of~(\ref{equ:S-ell-small},\ref{equ:S-ell-large}), where the density $\rho$ is obtained as $\langle\tilde{n}_{it}\rangle$
and the rod length $m=1+2^{b-1}$ (or $m=1$ for $b=0$).}
\label{fig:ent}
\end{figure}

We now present numerical results that illustrate how the patch repetition analysis is effective
in identifying and characterising metastable states in the East model, and how the results of this analysis can be used to 
predict dynamical correlation functions.  These results illustrate the operation of the scheme,
and the central role played by the averaging time $t$.  The implications of these results for other glassy systems
will be discussed in the following Section.

In the East model, we have evaluated $S_{\ell,t}$ numerically for the times $t$ shown with arrows in Fig.~\ref{fig:cor}b.
Fig.~\ref{fig:ent} shows our results (symbols), which we present by plotting $S_{\ell+1,t} - S_{\ell,t}$ as a function of $\ell$.   
We first observe that for $t\ll 1$ (i.e., $b=0$),
then $S_{\ell+1,t} - S_{\ell,t}$ is independent of $\ell$.  This result follows immediately from the trivial equilibrium
distribution $p^0(\CC)$ in the East model.  However, on averaging over larger time scales, structure appears in these
entropy measurements, and one may easily identify
a length scale $\ell^*$ associated with convergence of $S_{\ell+1,t} - S_{\ell,t}$ to its large-$\ell$ limit ${\cal S}_t$. 

To account for the $\ell$-dependence of $S_{\ell,t}$, recall that  the ``superdomain'' argument of~\cite{SE} 
indicates that sites with $\tilde{n}_{it}=1$
are separated by at least $2^{b-1}$ lattice sites (for $b\geq 1$).
This rule indicates that the statistics of the time-averaged profiles $\tilde{\CC}_t$ are related
to those of systems of hard rods of length $m=1+2^{b-1}$.  In fact, for times much less than $\tau_0$, we find that the statistics
of the profiles $\tilde{\CC}_t$ are almost exactly those of an ideal gas of hard rods of length $m$, where the leftmost
site of each hard rod carries a `1' and all others site carry a `0'.
If the number density of rods is $\rho$, straightforward counting arguments 
show that the patch entropies for this hard rod system
are given by 
\begin{equation}
S_{\ell}(m,\rho) =
-\ell\rho\log\rho -(1-\ell\rho) \log(1-\ell\rho), \quad  \ell \leq m, 
\label{equ:S-ell-small}
\end{equation}
and
\begin{equation}
S_{\ell}(m,\rho) =
S_m(m,\rho) + (\ell-m) {\cal S}^{\rm rod}(m,\rho) , \quad \ell > m,
\label{equ:S-ell-large}
\end{equation}
where 
\begin{multline}
{\cal S}^{\rm rod}(m,\rho) = -\rho\log\rho - (1-m\rho)\log(1-m\rho) \\ + [1-(m-1)\rho] \log[ 1-(m-1)\rho]
\label{equ:stot}
\end{multline}
 is the total entropy density.

The predictions of (\ref{equ:S-ell-small}) and (\ref{equ:S-ell-large}) are
shown in Fig.~\ref{fig:ent} as solid lines.  The fit is excellent.  The value of $\rho$ 
has been calculated from the numerical data (as $\rho=\langle \tilde{n}_{it}\rangle)$, but no 
other fit parameters are required (we take $m=1+2^{b-1}$).  We conclude that metastable states in the East model with lifetimes $t\sim c^{-b}$
are in one-to-one correspondence with configurations of hard rods of length $1+2^{b-1}$.  At equilibrium, states with equal numbers of rods are equiprobable, and the number density
of such hard rods is $\rho\approx c$. (To be precise, for $t\ll t_0$ then
$\rho=c-O(c^2)$: one has $\langle\overline{n}_{it}\rangle=c$, but the coarse-graining procedure leading to $\tilde{n}_{it}$ means that
the average number of sites with $\tilde{n}_{it}=1$ is less than the average number of sites with $n_i=1$.)   
It is however useful to recall at this point
that we have always been restricting our analysis to the situation $c^{-b} \ll \tau_0$, which also implies that $m\rho\ll 1$ (recall
$\tau_0$ is the bulk relaxation time).  
For times of order $\tau_0$, the situation is more complicated, since time scales are no longer well-separated~\cite{martinelli-arxiv}.  
To the the extent that that metastable states exist on these time scales, we expect them
still to be in one-to-one correspondence with hard-rod configurations, but states with equal numbers of rods are no longer equiprobable.  
The same limitations mean that the superdomain analysis of~\cite{SE} does not give quantitative results for structural
relaxation at equilibrium.

\begin{figure}
\includegraphics[width=8cm]{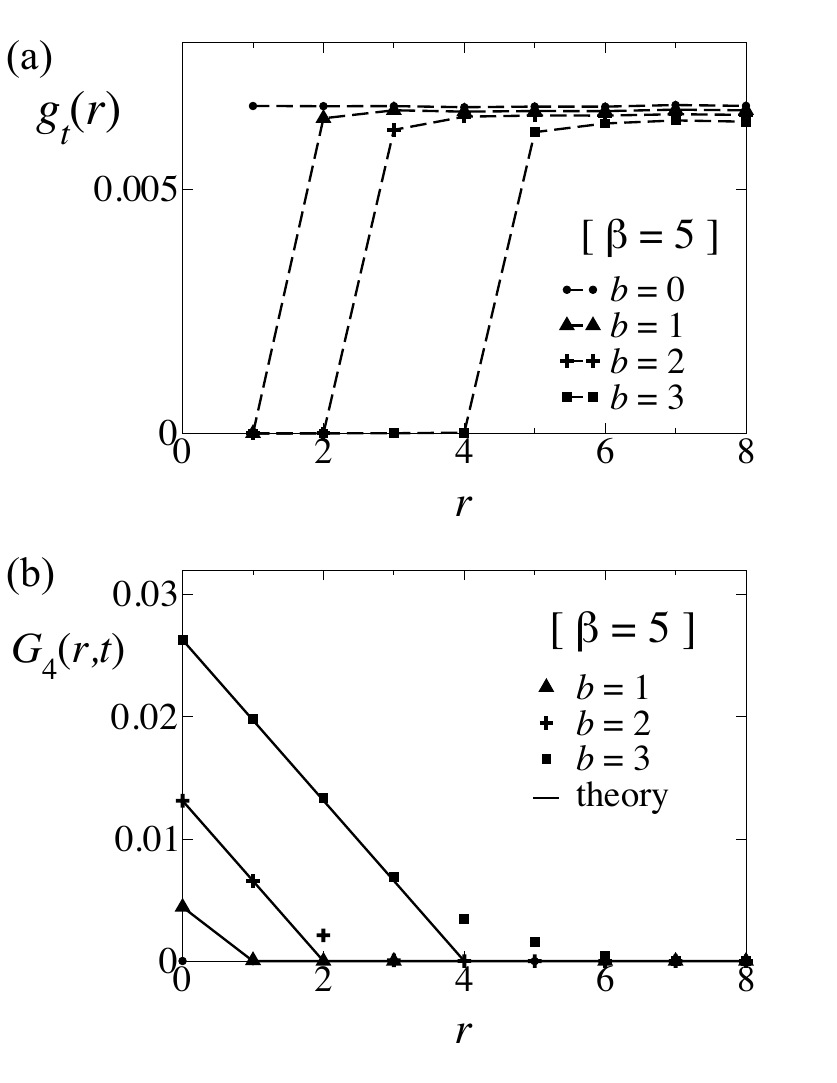}
\caption{Spatial correlation functions at $\beta=5$ and times indicated in Fig.~\ref{fig:cor}b.  (a) Two-point correlations $g_t(r)$ 
of the time-averaged density $\overline{n}_{it}$.  The dashed lines are guides to the eye.
The mapping to hard rods indicates that $g_t(r)$ should be equal to $0$ for 
$r\leq2^{b-1}$ and constant ($=\rho$) otherwise, consistent with the data.
(b) Two-point correlations $G_4(r,t)$ of the persistence 
function $p_{it}$.  The `theory' lines are predictions of~(\ref{equ:g4-rod}), where $\rho$ is fitted
to the value $G_4(0,t)$. }
\label{fig:gr}
\end{figure}

Fig.~\ref{fig:gr} shows the correlation function between time-averaged
densities 
\begin{equation}
g_t(r) = \frac{ \langle \overline{n}_{it} \overline{n}_{i+r,t} \rangle }{ \langle \overline{n}_{it} \rangle }.
\end{equation}
Given that the distribution over metastable states is that
of uncorrelated hard rods, one expects that $g_t(r)\approx0$ for $1\leq r\leq 2^{b-1}$ and $g_t(r)\approx\rho$ for $r>2^{b-1}$. 
This is confirmed numerically in Fig.~\ref{fig:gr}(a). 

Fig.~\ref{fig:gr}(b) shows the correlation between single-site ``persistences'' 
\begin{equation}
G_4(r,t) = \langle p_{it} p_{i+r,t} \rangle - \langle p_{it} \rangle^2,
\end{equation}
where the persistence $p_{it}=1$ if spin $i$ does not flip at all between
times $0$ and $t$; otherwise $p_{it}=0$.   Based on the superspin picture, we assume that relaxation within states 
is the only motion possible on times less than $t$, so
all sites should have $p_{it}=1$, except that each up spin in $\tilde{\CC}_t$ is followed
by $2^{b-1}$ spins with $p_{it}=0$.  

Assuming that this picture holds, one arrives (for $c^{1-b} \ll t \ll c^{-b}$) at
\begin{equation}
G_4(r,t) = {\rm max}[\rho( 1-r/2^{b-1}),0]
\label{equ:g4-rod}
\end{equation}
where $\rho$ is the number density
of hard rods discussed above.  
Comparing this prediction with the numerical data of Fig.~\ref{fig:gr}b, the agreement is reasonable, but there are significant deviations.
The data for $G_4(r,t)$ indicate that blocks of more than $2^{b-1}$ spins with $p_{it}=0$ occur more often than expected.  
This occurs because the persistence
$p_{it}$ is sensitive to (atypical) flips of spin $i$ that may occur even if this spin is not within a mobile
domain.  Such flips tend to occur to the right of mobile domains, increasing their apparent size.
On the other hand, the time-averaged density $\overline{n}_{it}$ is less sensitive to such rare spin flips: a significant
deviation of $\overline{n}_{it}$ from its typical behaviour requires that $n_i$ deviates from its typical behaviour for a time
comparable with $t$. 
This emphasises the effectiveness of the time-averaged spin $\overline{n}_{it}$ as a good measurement for probing
metastable states, while the persistence $p_{it}$ is less effective for this.

\section{Discussion of metastable states}
\label{sec:discuss}

\subsection{Length scales}

A central aim of~\cite{KL} was to identify a length scale (or perhaps several length scales) 
associated with ``amorphous order''.
It is clear from Figs.~\ref{fig:ent} and~\ref{fig:gr} that for times $t\sim c^{-b}$, there is a length scale 
$\ell^* \simeq 2^{b-1}$ associated
with the metastable states in the East model.
In general, this length scale should reflect structural features
associated with metastable states of lifetime $t$.  Here, this is simply the minimal possible
spacing between spins with $\tilde{n}_{it}=1$.  The longer-lived the states, the larger is $\ell^*$, and
the greater the degree of internal structure within the states. 

While these states can only be obtained by a dynamical construction (the time
averaging of the spin profile), we note that they
do have static structure, as measured (for example) by $g_t(r)$. 
Hence, one may explain properties of the system by a `free energy landscape' metaphor, of activated hopping
amongst states (basins), with each state having a distinct local structure.
The (time-dependent) length $\ell^*$ also matches the length
scale $\xi_4$ associated with dynamical heterogeneity in the model, which is usually measured (in the East model)
by the persistence correlation function $G_4(r,t)$.  The general inference here that is that the time-averaging
procedure used to determine $\ell^*$ will lead to coupling between $\ell^*$ and dynamical heterogeneities
associated with intra-state (relatively fast) dynamical motion.

We note that the length scale $\ell^*$ is obtained by considering the approach
of $S_{\ell+1,t}-S_{\ell,t}$ to its large-$\ell$ limit ${\cal S}_t$: an alternative route~\cite{KL,camma-patch}
is to expand the entropy for large $\ell$ as
$S_\ell = {\cal S}\ell^d[ 1 + (\xi/\ell)^{\nu-d} + \dots]$ where $d$ is the spatial dimension, $\xi$ a characteristic
length scale, and $\nu$ a scaling exponent.  Calculating results for the East model based on the hard rod
analysis, we have $d=1$ and $\nu=0$: the resulting
length scale is $\xi = \ell^* (\ell^* S_{\ell=1}- S_{\ell^*})/(2S_{\ell^*})$.
As may be inferred from Fig.~\ref{fig:ent}, the entropy
difference $\ell^* S_{\ell=1}-S_{\ell=\ell^*}$, is small compared to 
$S_{\ell^*}$, meaning that $\xi\ll\ell^*$.  The interpretation of $\xi$ is therefore not clear in this case:
it would be more appropriate to write $S_\ell = {\cal S}\ell^d[ 1 + a(\xi/\ell)^{\nu-d} + \dots]$ where $a$
is an amplitude (small in this case), and $\xi\approx \ell^*$ the true length scale.  However, extraction
the determination of two parameters $a$ and $\xi$ makes this method non-trivial [in this case
one might take $a=(S_{\ell=1}/{\cal S})-1$ but it is not clear that this is the best choice in general].

Another length scale~\cite{KL} that can be extracted from $S_\ell$ is the length $\ell_1$ for which $S_\ell\approx 1$.  In this
model, $\ell_1\sim 1/|c\log c|$, comparable (but not equal to) the typical distance $(1/c)$ between
up spins.  However, the length scale $\ell_1$ 
does not play any obvious role in the behaviour of the system.  A similar
conclusion was found in~\cite{camma-patch}.  However, contrary to the models considered in~\cite{camma-patch}, 
we find that $S_{\ell}/\ell$ approaches its large-$\ell$ limit from above in this model, and not from below.  
We do not have any simple physical argument for this observation, although we do note that the total entropy of the East
is very small at low temperatures, since the model does not account for the diversity of possible ``inactive'' states
that are commonly found in glassy systems~\cite{BBT,GC-bbt}.  

The conclusion of this analysis is that the $\ell$-dependence of $S_{\ell}$ does encode considerable information
about the range of correlations an amorphous system.  For the East model, it is relatively simple to identify the 
relevant physical length scale as $\ell^*$, but more generally, it remains unclear how to interpret measurements of
$S_{\ell}$.  Certainly, the subtleties associated with this interpretation should be borne in mind in future studies. 

\subsection{Atypical states and the complexity $\Sigma(f)$}

We established in Sec.~\ref{sec:results} that the  metastable states in the East model are in correspondence with
configurations of hard rods.  At equilibrium, the system typically occupies states whose number density
of rods is $\rho\sim c$.  However, as discussed in~\cite{KL}, it may be useful to consider ensembles in which
the model is biased into atypical metastable states (for example, with higher or lower values of $\rho$).

We define a free energy density for state $\alpha$ by writing the probability that the system is in that state as
\begin{equation}
P(\alpha) = \ee^{-\beta L f_\alpha}/Z ,
\label{equ:p-alpha}
\end{equation}
where $Z$ is the usual equilibrium partition function of the system.  
Based on the mapping to hard rods, we have
$f_\alpha = -\mu {\cal N}_\alpha$ where $\mu$ is a (negative) chemical potential, and
${\cal N}_\alpha$ is the number density of up spins in $\tilde{\CC}^\alpha$ (i.e., the number density of rods).  
For a given average density $\rho$,
the chemical potential $\mu$ may be obtained as $\beta \mu = -\frac{\partial}{\partial \rho} {\cal S}^{\rm rod}(m,\rho)$.

Now consider a modified ensemble in which states 
 occur with probability
\begin{equation}
P_q(\alpha) = \ee^{-\beta qLf_\alpha} / {\cal Z}(q)
\label{equ:q-ens}
\end{equation}
with ${\cal Z}(q) = \sum_\alpha  \ee^{-\beta q Lf_\alpha}$.  The ensemble with $q=1$ corresponds to equilibrium:
for large $q$ then states with more rods are suppressed while for $q<1$ they are enhanced.  As $q\to0$, all states
are equiprobable so the ensemble is dominated by the most numerous states (those with the highest entropy).

Since the probability of state $\alpha$ depends on $q$ through the exponential of an extensive quantity, the 
modified ensemble of (\ref{equ:q-ens}) is dominated by states that are very rare at equilibrium: such ensembles are the subject
of large deviation theory~\cite{touchette-review}.  To investigate the properties of this ensemble, it is useful to consider the average
free energy of states within the modified ensemble: 
\begin{equation}
f(q) = \sum_\alpha f_\alpha P_q(\alpha)=-\frac{1}{\beta L}\frac{\partial}{\partial q} \log {\cal Z}(q).
\end{equation}
Now rewrite the definition of ${\cal Z}(q)$ as
\begin{equation}
{\cal Z}(q) = \int\mathrm{d}f\, \ee^{\Sigma(f) -\beta qLf}
\label{equ:Zq-int}
\end{equation}
where (by definition) $\ee^{\Sigma(f)}$ is the density
of states with free energy $f$, so $\Sigma(f)$ is the (extensive) complexity~\cite{cav-review,KL}.  
For the East model, recall that the free energy $f_\alpha=-\mu{\cal N}_\alpha$,
and the density of states with a given number of rods is $\ee^{L{\cal S}^\mathrm{rod}(m,\rho)}$ [recall (\ref{equ:stot})].  Thus,
for the East model, we have a concrete expression for the complexity:
\begin{equation}
\Sigma_t(f) = L\,{\cal S}^{\rm rod}(m_t,-f/\mu_t),
\end{equation}
where $\mu_t$ is the rod chemical potential and $m_t$ is the relevant rod length; we have reintroduced
the subscript $t$ as a reminder that we are counting metastable states with lifetime much greater 
than some fixed time $t$, and that the rod size $m$ and chemical potential $\mu$ depend on this time.

It follows from (\ref{equ:Zq-int}) that the modified ensemble of (\ref{equ:q-ens}) maps to an ensemble of hard rod configurations
whose chemical potential is $q\mu$ (here $\mu<0$ is the rod chemical potential of the equilibrium ensemble with $q=1$).
As $q$ is reduced (the chemical potential becomes less negative), 
the number of hard rods increases.  As $q\to0$, one finds the maximum entropy state, $\rho=1/(m+1)$; continuing
to negative $q$, the system approaches the maximal density state $\rho=1/m$.
For $q>1$, the number of rods decreases, with $\rho\to0$ (and therefore $f\to0$) as $q\to\infty$.
The large-$q$ limit is similar to the entropy crises found in mean-field models~\cite{cav-review}, 
except that $\Sigma(f)\to0$ with a diverging gradient $d\Sigma/df\to\infty$: this is the reason that the transition
takes place as $q\to\infty$~\cite{KL}.  In summary, we find that $f(q)$ is a smooth function of $q$, with singular behaviour only
as $q\to\pm\infty$.  This means that phase transitions only occur at zero temperature, as expected in the East model~\cite{Ritort-Sollich,GST-kcm}.

\subsection{Biased ensembles based on patches and the role of dynamical fluctuations}

In the previous subsection, we analysed atypical states in the East model using the superspin picture of~\cite{SE} and
the numerical results that we obtained for typical states.
Kurchan and Levine~\cite{KL} proposed a method for numerical investigation of atypical states, via the 
ensemble (\ref{equ:q-ens}).
However, numerical sampling of such biased ensembles is difficult,
since these ensembles are dominated by configurations that are far from typical.  In the remainder
of this Section we analyse different methods for analysing rare metastable states.  We use the East model
as a representative example, but the main aspects of the discussion apply quite generally to models
with many metastable states.

The method proposed in~\cite{KL} for analysis of atypical states
was based on Renyi complexities (see for example~\cite{paladin86}): 
the idea is that one estimates a quantity $Y_{{\cal B}}$ that can be used to infer
the free energy associated
with a patch $\cal B$ of size $\ell$.   Each possible patch has a free energy density 
that is estimated as $Y_{\cal B}=(-1/\ell) \log [n({\cal B})/L]$
where $n({\cal B})$ is the number of times that patch $\cal B$ appears in $\tilde{\CC}_t$ [recall (\ref{equ:S-patch})].  
Writing $n({\cal B})/L = \ee^{-\ell Y_{\cal B}}$
and comparing with (\ref{equ:p-alpha}) 
motivates the analogy between $Y_{\cal B}$ and the free energy density $\beta f_\alpha$.

Then, one may analyse the $q$-ensemble by giving extra statistical weight to patches
with large (or small) values of $Y$.  That is, consider a patch-analogue of (\ref{equ:q-ens}) where one assigns
patch probabilities:
\begin{equation}
p_q({\cal B}) = \frac{ \ee^{-q \ell Y_{\cal B}} }{ {\cal Z}_{\rm p}(q) }
\label{equ:q-patch}
\end{equation}
with ${\cal Z}_{\rm p}(q)=\sum_{\cal B} \ee^{-q \ell Y_{\cal B}}$.  Since $Y_{\cal B}$ can be estimated directly via $n({\cal B})$,
one may also estimate a free-energy-like quantity, which is the Renyi complexity:
\begin{align}
K_{q,\ell} 
&
= \frac{1}{\ell(q-1)} \log \sum_{\cal B} \ee^{-q \ell Y_{\cal B}} 
\nonumber \\ &
 = \frac{1}{\ell(q-1)} \log \sum_{\cal B} [n({\cal B})/L]^q.
 \label{equ:renyi}
\end{align}  
The patch quantity $(q-1)K_{q,\ell}$ is analogous to
the quantity $(1/L)\log {\cal Z}(q)$, defined for states in the previous section.
For large enough $\ell$, one expects the modified ensemble defined by (\ref{equ:q-patch}) to resemble that defined by (\ref{equ:q-ens}).

From a numerical perspective, this approach is difficult -- one is attempting to reconstruct a 
biased ensemble using data from an equilibrium one, and typical configurations from the two ensembles are quite different 
from one another.  The usual approach to this problem is to use a biased sampling scheme such as umbrella sampling~\cite{frenkel-smit}
for ensembles of configurations, or transition path sampling~\cite{TPS-review} for ensembles of trajectories.  Here the situation
is subtle: while the bias in (\ref{equ:q-ens}) appears to be a configurational one, based on free energies of states,
we recall that any practical sampling scheme uses \emph{time-averaged} profiles to infer the relevant states.  
This means that patches that are rare in (\ref{equ:renyi}) may be associated with unusual metastable states,
or they may be associated with rare dynamical events in which the averaged profile $\overline{\CC}_t$ does
not converge to any of the profiles $\overline{\CC}_t^\alpha$ that are associated with metastable states in the
system.  This motivates
us to consider the relation to dynamical sampling methods and recent work on dynamical large deviations.

\subsection{Relation to dynamical large deviations}

The statistical properties of \emph{time-integrated} dynamical quantities have received considerable
interest recently, especially through studies of large deviations in glassy systems~\cite{jack-rom09,hedges09,elmatad10,jack11-stable,Speck12,Merolle,kcm-transition,elmatad13} 
(including the East model~\cite{Merolle,kcm-transition,elmatad13}).
The most
relevant situation for this work is the one considered for a spin-glass model in \cite{jack-rom09}, where
the time $t$ (there called $t_{\rm obs}$) is well-separated from both a fast (intra-state)
relaxation time scale $t_{\rm f}$ and a slow (inter-state) relaxation time $t_{\rm s}$.   
The question of interest there is: given a quantity $k(\CC)$ that can be evaluated for a given
configuration, what is the probability
distribution of $K = \int_0^t k(\CC(t)) dt$?  Clearly, this is closely related to distributions
of the quantities
considered here, such as $\overline{n}_{it} = (1/t) \int_0^t n_i(t')dt'$.
We note that in~\cite{jack-rom09}, the \emph{fast} time
scale $t_{\rm f}$ was comparable with the equilibrium relaxation time of the system, while the \emph{slow} time scale
was much longer still.  Here we consider the case where $t_{\rm f} \ll t \ll t_{\rm s} \ll \tau_0$: for example
$t_{\rm f}=c^{1-b}$ and $t_{\rm s}=c^{-b}$, as above.  Nevertheless, the main results of~\cite{jack-rom09} apply also in this case.

In particular, one may consider a biased ensemble where a dynamical trajectory $\CC(t)$ occurs with probability
\begin{equation}
\mathrm{Prob}[\CC(t)|s] = \mathrm{Prob}[\CC(t)|0] \cdot \ee^{-sK[\CC(t)]} \cdot \frac{1}{ Z_\mathrm{d}(s) } .
\label{equ:s-ens}
\end{equation}
Here, the short-hand
notation $[\CC(t)]$ indicates dependence on a trajectory of the system (with time running from $0$ to $t$), 
while $\mathrm{Prob}[\CC(t)|0]$ is the probability of trajectory $\CC(t)$ at equilibrium, and $Z_\mathrm{d}(s)$ 
is a normalisation constant.   Following~\cite{jack-rom09}, in the joint limit $t_{\rm f} \ll t \ll t_{\rm s}$, this biased ensemble
develops a singular dependence on $s$.  
That is, if the average value of $k$ within state $\alpha$ is $k_\alpha$ then for $s>0$
the ensemble is dominated by the metastable state(s) with minimal $k_\alpha$, while for $s<0$ 
it is dominated by state(s) with maximal $k_\alpha$.

If one replaces $s\to\lambda/t$ in (\ref{equ:s-ens}), one sees that trajectories are now being biased according to their average
values of $k(\CC)$ (instead of their integrated value).  Assuming that dynamical fluctuations can be neglected 
(that is, $\overline{\CC}_t\approx
\overline{\CC}^\alpha$, due to separation of time scales), and that every trajectory is localised in a single metastable state,
one arrives at
\begin{equation}
\mathrm{Prob}[\CC(t)|s] \approx \mathrm{Prob}[\CC(t)|0] \cdot \ee^{-\lambda k_{\alpha[\CC(t)]}} \cdot \frac{1}{Z_\mathrm{d}(\lambda/t)}
\end{equation}
where $\alpha[\CC(t)]$ is the state within which $\CC(t)$ is localised.  
In that case, one may write the probability of state $\alpha$ as
\begin{equation}
P_\lambda(\alpha) \approx \ee^{-\beta L f_\alpha} \cdot \ee^{-\lambda k_{\alpha}} \cdot \frac{1}{ Z_{\rm s}(\lambda) }
\label{equ:lambda-ens}
\end{equation}
where $Z_{\rm s}(\lambda)$ is a normalisation constant.
Physically, (\ref{equ:lambda-ens}) indicates that the effect of weak bias 
$s=O(1/t)$ in (\ref{equ:s-ens}) is to reshuffle probability between metastable states, in a similar
way to that anticipated in (\ref{equ:q-ens}).  One may also conjecture that
the singular dependence of the ensemble in (\ref{equ:s-ens}) on the parameter $s$ might be 
resolved as a smooth change on a scale $s=O(1/t)$.  (Note however that this behaviour would still be far beyond 
any linear-response regime.)

Further, if $k(\CC)$ can be chosen so that $k_\alpha \approx \beta L f_\alpha$, then (\ref{equ:lambda-ens}) 
reduces to (\ref{equ:q-ens}), with $\lambda=q-1$.
That would allow the ensemble of (\ref{equ:q-ens}) to be sampled via the numerical methods that have been used already to sample
the $s$-ensemble.  In fact this is a relatively simple matter in the East model,
since $f_\alpha=-\mu{\cal N}_\alpha$ is tightly correlated with the total density of up spins $L^{-1}\sum_i n_i$, whose
large deviations were considered in~\cite{Merolle}.
However, the construction required to obtain the statistics of metastable states  is different from that used in most 
$s$-ensemble studies so far, since it would require analysis of trajectories with $1\ll t \ll \tau_{\rm eq}$ using $s=O(1/t)$  
(where $\tau_{\rm eq}$ is the structural relaxation time): previous
studies concentrated on the alternative limit $s=O(1)$ and $t\gg \tau_{\rm eq}$. The role of dynamical fluctuations
also needs careful consideration when comparing results: we are assuming here that $\overline{\cal C}_t$ is always close
to the relevant $\overline{\CC}^\alpha$ but ensuring that this is indeed the case for trajectories in the $s$-ensemble
requires careful examination of the limit $t_{\rm f} \ll t \ll t_{\rm s}$.

Finally, we recall that for non-zero $c$, the East model
has a phase transition in the $s$-ensemble, in the limit where $t\gg \tau_0$~\cite{kcm-transition}. 
(This is distinct from possible singular
behaviour if $c^{1-b} \ll t \ll c^{-b}$ which clearly requires $c\to0$.)  It is clear from~\cite{Merolle,jack-spacetime06}
that this transition is associated with a ``phase-separated'' regime where time-averaged profiles become macroscopically
inhomogeneous, containing a large ``bubble'', free of excitations.  In the language of hard rods, this bubble corresponds
to a large rod (length $m$ of the same order as the system size), with a commensurately long lifetime (diverging
with the system size).  In general, if states with very long lifetimes exist, one expects them to dominate the system 
for $s>0$ and $t\to\infty$: the necessary link~\cite{KL} between the very long lifetime and a diverging length scale
is particularly clear KCMs such as the East model~\cite{kcm-transition}.

\section{Conclusion and outlook}
\label{sec:conc}

We have shown
that the patch-repetition analysis 
of Kurchan and Levine~\cite{KL} is useful in identifying metastable states in the East model, and analysing their structure.
The results are consistent with the analysis of Sollich and Evans~\cite{SE}, 
as shown by the excellent fits in Fig.~\ref{fig:ent}.  Knowledge of the metastable
states can also be used to predict dynamical correlation functions as shown in Fig.~\ref{fig:gr}.

We have emphasised that the patch-repetition analysis is inherently dynamical in nature, because of its construction
from time-averaged density profiles.  In this sense, it can be used to identify metastable states even in systems
where the potential energy landscape is flat and featureless.  (This observation can also be rationalised through the idea that
the flat landscape is endowed with a non-trivial metric~\cite{whitelam-metric}, 
encapsulating the idea that motion within some regions of the landscape
may be much faster than motion between those regions.)  

Finally, we compared several different ways of analysing the distribution of metastable states within a system, through biased
ensembles in which patch probabilities are reweighted (\ref{equ:q-patch}) and through ensembles where trajectories are biased by some
time-averaged measure (\ref{equ:s-ens}).  In the end, these methods for numerical analysis of metastable states must be tested on atomistic
models, to understand how efficient they are and how useful their results will be.  We hope that future studies in this area
will be forthcoming.

\acknowledgments
I would like to thank Jorge Kurchan, Peter Sollich, and Giulio Biroli, for enlightening discussions on metastable states,
patches and large deviations.  This work was funded by the EPSRC through grant EP/I003797/1.

\end{document}